
\magnification=1200  
%
\def\newline{\hfill\penalty -10000}  
\def\title #1{\centerline{\rm #1}}
\def\author #1; #2;{\line{} \centerline{#1}\smallskip\centerline{#2}}
\def\abstract #1{\line{} \centerline{ABSTRACT} \line{} #1}
\def\heading #1{\line{}\smallskip \goodbreak \centerline{#1} \line{}}
\newcount\refno \refno=1
\def\refjour #1#2#3#4#5{\noindent    \hangindent=1pc \hangafter=1
 \the\refno. #1, #2 ${\bf #3}$, #4 (#5). \global\advance\refno by 1\par}
\def\refbookp #1#2#3#4#5{\noindent \hangindent=1pc \hangafter=1
 \the\refno.  #1, #2 (#3, #4), p.~#5.    \global\advance\refno by 1\par}
\def\refbook #1#2#3#4{\noindent      \hangindent=1pc \hangafter=1
 \the\refno.  #1, #2 (#3, #4).           \global\advance\refno by 1\par}
\newcount\equatno \equatno=1
\def\adveqn{(\the\equatno) \global\advance\equatno by 1}

\def\up#1{\leavevmode \raise 0.2ex\hbox{#1}}

\newcount\figno
\figno=0
\def\figure{\global\advance\figno by 1 Figure~\the\figno.~}
%

%
\vsize=8.75truein
\hsize=5.75truein
\hoffset=0.5truein
%
\baselineskip=0.166666truein
%
\parindent=25pt
%
\parskip=0pt
%


\def\sles{\lower2pt\hbox{$\buildrel {\scriptstyle <}
   \over {\scriptstyle\sim}$}}
\def\sgreat{\lower2pt\hbox{$\buildrel {\scriptstyle >}
   \over {\scriptstyle\sim}$}}
\line{}
\title{\bf GAMMA-RAY BURSTS FROM  NEUTRON STAR MERGERS}
\author
Tsvi Piran\parindent=0pt  ;
Racah Institute for Physics, The Hebrew University, Jerusalem 91904, Israel ;

\abstract{ Binary neutron stars merger (NS$^2$M) at cosmological
distances is probably the only $\gamma$-ray bursts model based on an
independently observed phenomenon which is known to be taking place at
a comparable rate. We describe this model, its predictions and some
open questions.  }

\heading{Cosmological $\gamma$-Ray Bursts and Fireballs}

Compton-GRO has demonstrated, quite convincingly, that $\gamma$-ray
bursts (grbs) originate from cosmological sources$^{1,2}$.  Evidence
for the predicted$^{3,4}$ correlations between the duration, the
strength and the hardness of the bursts begins to emerge$^{5,6}$.
Preliminary analysis suggests that the weakest bursts originate from
$z \approx 1$, in agreement with a cosmological $C/C_{min}$
distribution$^{4}$, corresponding to a local rate of
$\approx 10^{-6}/{\rm year}/{\rm galaxy}$ (depending on the
cosmological model and on other factors).  The energy released in each
burst depends also on the cosmological model $ 10^{50} \sgreat E \sles
10^{51}$ergs if the energy emission is isotropic.

The intense energy released in a small volume (evident by the rapid
rise time of some of the  pulses) implies that any cosmological grb
source is initially optically thick$^{7}$ to $\gamma\gamma \rightarrow
e^+ e^-$.  The large initial optical depth prevent us from observing
directly the photons released by the source regardless of the specific
nature or the source.  The sources produce an optically thick
radiation-electron-positrons plasma ``fireball", which  behaves like a
fluid, expands and reaches relativistic velocities$^{8,9}$.  The
observed radiation emerges only after the fireball has expanded
significantly and became optically thin.

One should divide, therefore, the discussion of cosmological grbs to a
discussion of the nature of the energy source (for which we present a
model here) and a discussion of the fireball phase (which we address
elsewhere in this volume).  For the paper  it is sufficient to recall
that the fireball must reach ultra-relativistic velocities with a
Lorentz factor $\gamma \sgreat 10^2$ to produce a grb. Since $\gamma
\approx E/Mc^2$ (where E is the total energy of the fireball and M is
the mass of the baryons in the fireball) the condition $\gamma > 10^2$
sets a strong upper limit on the amount of baryons: $M < E/\gamma c^2
\approx .5 10^{-5} M_\odot (E/10^{51} ergs) (\gamma/10^2)^{-1}$. This
condition poses a strong constraint on grb models.


\heading{NS$^2$M and GRBs - Agreement at a Glance}

Neutron star binaries, such as the one observed in the famous binary
pulsar PSR 1916+13, end their life in a catastrophic merge event
(denoted here NS$^2$M).  Using the three observed binary pulsars we
can estimate the expected rate of NS$^2$M events$^{10,11}$ as $\approx
10^{-5.5 \pm .5}/{\rm year}/{\rm galaxy}$.  An energy comparable to a
neutron star binding energy ($\sgreat 5 \times 10^{53}$ ergs) is
released in NS$^2$Ms mostly as neutrinos  and gravitational radiation.
The neutrino signal is comparable in its signature to supernova
neutrino signals which are thousand times more frequent. It is
unlikely that it will ever be detected. The gravitational radiation
pulses, have however, a unique signature and Two gravitational
radiation detectors, LIGO and VIRGO are currently constructed to
detect them.

Several years ago Eichler, Livio Piran and Schramm$^{13}$, (see also
$^{14,15,16,17}$) suggested that grbs originate at NS$^2$Ms.
Between $10^{-2}$ to $10^{-3}$ of the total energy released in NS$^2$Ms
is sufficient to power a grb at a cosmological distance. The required
energy could be converted to electromagnetic energy either
via${13,15}$ $\nu \bar\nu \rightarrow e^+ e^-$ or via magnetic
processes in an accretion disk that forms in the merger$^{18}$.  The
rates of NS$^2$Ms estimated from binary pulsars and the observed rate
of grbs measured by BATSE and estimated from cosmological fits are
within half an order of magnitude from each other. A remarkable
agreements in view of the large uncertainties involved in both
estimate.

\heading{Numerical Simulations of NS$^2$M - Some Answers to Further Questions}

It worthwhile, therefore,  to explore whether the mergers can produce
clean enough fireballs (i.e.  fireballs with sufficiently low baryonic
load) as required from the fireball analysis and to ask whether enough
energy can be converted to electromagnetic energy in this events.
To address these iissues  we$^{19}$ developed a numerical code that
follows neutron star binary mergers and calculates the thermodynamic
conditions of the coalesced binary. The process of coalescence, from
initial contact to the formation of an axially symmetric object, takes
only a few orbital periods. Some of the material from the two neutron
stars is shed, forming a thick disk around the central, coalesced
object. The mass of this disk depends on the initial neutron star
spins; higher spin rates resulting in greater mass loss, and thus more
massive disks. For spin rates that are most likely to be applicable to
real systems, the central coalesced object has a mass of $2.4M_\odot$,
which is tantalizingly close to the maximum mass allowed by any
neutron star equation of state for an object that is supported in part
by rotation.  Using a realistic nuclear equation of state we estimate
the temperatures after the coalescence: the central
object is at a temperature of $\sim 10$MeV, whilst the disk is heated
by shocks to a temperature of 2-4MeV.

A typical density cut perpendicular to the equatorial plan is shown in
Fig. 1.  The disk is thick, almost toroidal; the material having
expanded on heating through shocks.  This disk surrounds a central
object that is somewhat flattened due to its rapid rotation.  An
almost empty centrifugal funnel forms around the rotating axis and
there is practically no material above the polar caps.  This funnel
provides a region in which a baryon free radiation-electron-position
plasma could form$^{20}$.  Neutrinos and antineutrinos from the disk
and form the polar caps would collide and annihilate preferentially in
the funnel (the energy in the c.m. frame is larger when the colliding
$\nu$ and $ \bar \nu$ approach at obtuse angle, a condition that
easily holds in the funnel). The numerical computations do not show
any baryons in the funnels. The resolution of our computation is
insufficient, howeer, to show that the baryonic load in the funnel
is as low as needed.  The neutrinos radiation pressure on polar cap
baryons can generate a baryonic wind that will load the flow.
Estimates of this effect$^{21,22}$ show that it is negligible if the
temperature on the polar caps is sufficiently low. The estimated
temperature from our computations is $\approx 2$MeV, which is
marginal. Our temperature estimate is, however, least certain in low
temperature regions like this.

If the core does not collapse directly to a black hole it will emit
its thermal energy as neutrinos. The neutrino flux is sufficiently
large that $\approx 10^{-2}$ to $10^{-3}$ of it could be converted to
electron-positron pairs via $\nu \bar \nu \rightarrow e^+ e^-$ and
produce a grb.  The time scale for the neutrino burst is short enough
to accommodate even the shortest rise times observed.  An additional
energy source that could power a grb is the accretion of the disk
surrounding the central object. This energy source can operate on a
longer time scale and it takes place regardless of the question of
whether the central object collapse directly to a black hole or not.

\heading{Open Questions and Predictions}

The numerical calculations support earlier suggestions$^{17}$ that
the energy release in anisotropic and that an empty funnel forms
around the rotating axis of the binary system.  The fireball is highly
non spherical and it expands along the polar axis and forms a jet.
This poses an immediate constraint on the model.  If the width of the
jet is $\theta$ than we observe grbs only from a fraction
$2\theta^{-2}$ of NS$^2$Ms. The rates of grbs and NS$^2$Ms agree
only if $\theta \sgreat 0.2$ (unless the rate of NS$^2$Ms is much
higher than the current estimates). A condition which at first glance
is satisfied by the funnel seen in Fig. 1.

The duration and spectra of grbs vary greatly from one burst to another.
Both are
determined by the fireball phase but the source might contribute in
producing fireballs with different Lorentz factors and different
initial durations.  Within
the funnel the  baryonic load will vary as a function of the
angular position leading to varying
final Lorentz factors which, in turn, produce bursts with different
durations and spectra.
Another
source of variability could arise from the interplay between the two
energy sources in NS$^2$Ms: Neutrino annihilation
and accretion energy of the disk. These  mechanisms would
operate on different time scale and produce different looking bursts.
An additional source of diversity$^{19}$ is the
distinction between systems that collapse directly to a black hole and
those that undergo a longer rotating core phase.
Finally, black hole-neutron star binaries are predicted to be as common
as neutron star binaries$^{10}$. A black hole neutron star merger$^{14}$
would produces grbs with different characteristics  than NS$^2$M.

NS$^2$M events can take place in a variety of host systems including
dwarf galaxies, or even in the intergalactic space if the neutron star
binary is ejected from the host galaxy when it forms$^{18}$.  Hence,
unlike other cosmological models it is not essential that an optical
counter part will be observed in the location of grbs$^{23}$.  A
unique prediction of the NS$^2$M model is that grbs should be
accompanied by gravitational radiation signals from the final stages
of the merger and vice versa (the latter is true only up to the
anisotropic emission factor discussed earlier). This coincidence could
prove or disprove this model. It could also serve to increase the
sensitivity of the gravitational radiation detectors$^{12}$.
Hopefully, this coincidence will be detected and the model will be
confirmed when gravitational radiation detectors will become
operational at the turn of the century.

I would like to thank Ramesh Narayan for many helpful discussions.
This work was supported in part by a BRF grant to the Hebrew
University and NASA grant NAGS-1904 to the CFA.

\heading{Reference}
\def\ApJ{{\it Ap. J.}}
\def\ApJL{{\it Ap. J. L.}}
\def\Nature{{\it Nature}}
\def\etalc{{\it et. al., }}
\item
{1.} Meegan, C.A., \etalc 1992, \Nature, {\bf 355} 143.
\item
{2.} Meegan, C.A., \etalc 1993, this volume.
\item
{3.} Paczy\'nski, B. 1992, \Nature, 355, 521.
\item
{4.} Piran, T., 1992, \ApJL {\bf 389}, L45.
\item
{5.} Norris \etalc 1993, this volume.
\item
{6.} Davis \etalc 1993, this volume.
\item
{7.} Piran, T. and Shemi, A.,  1993, \ApJL, {\bf 403}, L67.
\item
{8.} Goodman, J., 1986, \ApJL, {\bf 308} L47.
\item
{9.} Paczy\'nski, B., 1986, \ApJL, {\bf 308}, L51.
\item
{10.} Narayan, R., Piran, T. and Shemi, A., 1991, \ApJL, {\bf 379}, L17.
\item
{11.} Phinney, E. S., 1991, \ApJL, {\bf 380}, L17.
\item
{12.} Kochaneck C. and Piran, T., 1993, \ApJL, in press.
\item
{13.} Eichler, D., Livio, M., Piran, T., and Schramm, D. N.
1989, \Nature, {\bf 340}, 126.
\item
{14.} Paczy\'nski, B., 1991, Acta Astronomica, {\bf 41}, 257.
\item
{15.} Goodman, J., Dar, A. and Nussinov, S. 1987, \ApJL, {\bf 314}, L7.
\item
{16.} Piran, T., 1990, in Wheeler, J. C., Piran, T. and Weinberg, S.
{\it  Supernovae} World Scientific Publications.
\item
{17.} Piran, T., Narayan, R. and Shemi, A., 1992,
in Paciesas W. S. and  Fishman, G. J. eds.
{\it  Gamma-Ray Burst, Huntsville, 1991}, AIP press, 149.
\item
{18.} Narayan, R., Paczy\'nski, B., and Piran, T., 1992, \ApJL,
{\bf 395}, L83.
\item
{19.} Davies, M. B., Benz, W., Piran, T.,  and Thielemann, F. K. 1993,
submitted to \ApJ.
\item
{20.} Mochkovich, R. \etalc 1993, this volume.
\item
{21.} Duncan, R., Shapiro, S. L., and Wasserman, I., 1986, \ApJ,
{\bf 340}, 126.
\item
{22.} Woosley, S. E., and Baron, E., 1992, \ApJ, {\bf 391}, 228.
\item
{23.} Schaffer, B. \etalc 1993, this volume.
\bigskip
\heading{Figure Captions}
\item{Fig. 1}
Logarithmic density contour lines at the end of the computation of the
merger. The contours are logarithmic, at intervals of 0.25 dex (from
$^{19}$).
\end